\documentclass[structabstract]{aa}  

\usepackage[latin1]{inputenc} 		
\usepackage{gensymb} 			

\usepackage{natbib}
\bibpunct{(}{)}{;}{a}{}{,}
\usepackage{txfonts}

\usepackage{graphicx}
\usepackage{color}
\usepackage{url}

\def \isdc        {INTEGRAL Science Data Centre, Universit\'e de Gen\`eve, Chemin d'Ecogia 16, CH-1290
                  Versoix, Switzerland}

\def \obsgen      {Observatoire de Gen\`eve, Universit\'e de Gen\`eve, Chemin des Maillettes 51, CH-1290
                Sauverny, Switzerland}

\def \liege      {Institut d'Astrophysique et de G\'eophysique, Universit\'e de Li\`ege, All\'ee du 6-Ao\^ut 17, B\^atiment B5c, B-4000 Li\`ege, Belgium}

\begin{document}

\title{$\eta$ Carinae: a very large hadron collider}
\author{R. Walter\inst{1, 2}, C. Farnier\inst{1, 2}, J.-C. Leyder\inst{1,3}\fnmsep\thanks{FNRS Research Fellow} }
\offprints{Roland.Walter@unige.ch}

\institute{\isdc \and \obsgen \and \liege}

\date{Received $<date>$; Accepted $<date>$}

\abstract
{$\eta$ Carinae is the colliding wind binary with the largest mass loss rate in our Galaxy and the only one in which hard X-ray emission has been detected.}
{$\eta$ Carinae is therefore a primary candidate to search for particle acceleration by probing its gamma-ray emission. }
{We used the first 21 months of \textit{Fermi/LAT} data to extract gamma-ray ($0.2-100$ GeV) images, spectra and light-curves, then combined them with multi-wavelength observations to model the non-thermal spectral energy distribution.}
{
A bright gamma-ray source is detected at the position of $\eta$ Carinae. Its flux at a few 100 MeV corresponds very well to the extrapolation of the hard X-ray spectrum towards higher energies.
The spectral energy distribution features two distinct components. The first one extends over the keV to GeV energy range, and features an exponential cutoff at $\sim 1$~GeV. It can be understood as inverse Compton scattering of ultraviolet photons by electrons accelerated up to $\gamma\sim 10^4$ in the colliding wind region. 
The expected synchrotron emission is compatible with the existing upper limit on the non-thermal radio emission.
The second component is a hard gamma-ray tail detected above 20 GeV. It could be explained by $\pi^0$-decay of accelerated hadrons interacting with the dense stellar wind. The ratio between the fluxes of the $\pi^0$ and inverse Compton components is roughly as predicted by simulations of colliding wind binaries. This hard gamma-ray tail can only be understood if emitted close to the wind collision region. 
The energy transferred to the accelerated particles ($\sim5\%$ of the collision mechanical energy) is comparable to that of the thermal X-ray emission. 
}
{We have measured the electron spectrum responsible for the keV to GeV emission and detected an evidence of hadronic acceleration in $\eta$ Carinae. These observations are thus in good agreement with the colliding wind scenario suggested for $\eta$ Carinae.}
{}

\keywords{Gamma rays: stars -- X-rays: binaries -- X-rays: individuals: $\eta$ Carinae -- X-rays: individuals: FGL J1045.0-5942 -- Acceleration of Particles}

\maketitle

\section{Introduction}
\label{sec:Introduction}

About 30 early-type stellar systems feature synchrotron radiation in the radio domain, a signature of electron acceleration \citep{2007A&ARv..14..171D}. Diffusive shock acceleration in stellar wind collisions \citep{2003A&A...399.1121B}, either in colliding wind binaries or OB associations, is the most likely acceleration process and a candidate for cosmic ray acceleration \citep{1981NYASA.375..297A, 1982ApJ...258..860C}. Gamma-rays, emitted by hadrons accelerated in stellar wind collisions, have however not yet been 
identified.

The gamma-ray emission expected from a colliding wind binary increases with the stellar wind mechanical luminosity, the fraction of the wind enduring collision, the photon energy density for inverse Compton emission and the matter density for proton-proton interaction and subsequently $\pi^0$-decay. Located at a distance of 2.3~kpc \citep{Smith-06},
\object{$\eta$ Carinae} is one of the most massive (80--120~$M_{\sun}$; \citealt{Davidson+97, Hillier+01}) and brightest stellar system in the Galaxy and features the strongest mass loss rate known. As, in addition, its colliding wind region is relatively wide, $\eta$ Carinae is a primary candidate for gamma-ray detection.

The ``Great Eruption'' of 1843 saw $\eta$~Carinae become the second brightest object in the sky; it was followed by another noteworthy flare in 1890 \citep[see e.g.][]{Davidson+99,Fernandez-Lajus+10}.
The \textit{Homunculus} is the name given to the extended bipolar nebula observed around $\eta$~Carinae: it was crafted by the colossal amount of matter (10--20~$M_{\sun}$; \citealt{Smith+03}) that was ejected during the Great Eruption. The second burst led to the ejection of $\simeq 1$~$M_{\sun}$, creating the so-called \textit{little homunculus}. $\eta$ Carinae keeps emitting matter thanks to its powerful stellar winds : the mass-loss rate is believed to be $10^{-4}$--$10^{-3}$~$M_{\sun}$\,yr$^{-1}$ \citep{Andriesse+78, Hillier+01, 2002A&A...383..636P, vanBoekel+03}.

Even if not all questions are settled, strong evidences suggest that $\eta$~Carinae is a binary system. For instance, radio \citep{Duncan+95}, millimeter (mm; \citealt{Abraham+05}), optical \citep{Damineli-96, Damineli+00}, near-infrared (near-IR; \citealt{Whitelock+94, Whitelock+04, Damineli-96}), and X-ray \citep{Corcoran-05} observations, obtained over the last decades, have unveiled the existence of a period of $\sim2022.7 \pm 1.3$~days \citep{Damineli+08-periodicity}.

The first component in the binary system is very likely a luminous blue variable (LBV; \citealt{Davidson+97}). The second component is probably a late-type nitrogen-rich O or Wolf-Rayet (WR) star \citep{Iping+05, Verner+05}.
The semi-major axis of the orbit is $16.64$~AU \citep{Hillier+01}, and the eccentricity is very high ($e \sim 0.9$; \citealt{Nielsen+07}). Thus, the periastron distance $r_{\mathrm{periastron}}$ is around 1.66~AU, while the primary star's radius $R_{\star, 1}$, although poorly constrained, is estimated between 0.7--1 AU \citep[resp. ][]{Corcoran+07,Damineli-96}.

The X-ray emission of $\eta$~Carinae can be discriminated into two components, precisely known thanks to \textit{Chandra} observations \citep{Seward+01}. The softer part ($kT_{\mathrm{SX}} \sim 0.5$~keV) prevails in the spectrum up to 1.5~keV. It is spatially extended, and is likely linked to the stellar winds colliding with the interstellar matter. The harder part ($kT_{\mathrm{HX}} \sim 4.7$~keV) dominates over the 2--10~keV domain. It is punctual, centered on the binary system, and believed to be due to the hydrodynamical shock created by the stellar winds of both components colliding with each other.

The X-ray spectrum of $\eta$~Carinae supports the colliding-wind binary scenario \citep{Usov-92, Stevens+92, Corcoran-05, Pittard+07, 2009MNRAS.394.1758P}. The basic idea is that the LBV emits a rather slow and dense stellar wind ($v_{\infty, 1} \simeq 500$~km/s, $\dot{M}_{1} \simeq 2.5 \times 10^{-4}$~$M_{\sun}$/yr; \citealt{2002A&A...383..636P}), which collides into the faster, shallower wind originating from the secondary component ($v_{\infty, 2} \simeq 3000$~km/s, $\dot{M}_{2} \simeq 1.0 \times 10^{-5}$~$M_{\sun}$/yr; \citealt{2002A&A...383..636P}). In this framework, the wind collision region represents the location of the hydrodynamical shock, and thus of the X-ray emitting region.

A hard X-ray tail had first been observed towards $\eta$~Carinae by \textit{BeppoSAX} \citep{Viotti+04}, and has subsequently been unambiguously confirmed by both \textit{INTEGRAL} \citep{2008A&A...477L..29L, Leyder2010} and \textit{Suzaku} \citep{Sekiguchi+09} observations. This gives a strong evidence that the wind collision leads to non-thermal electron acceleration, and suggests the possibility of a gamma-ray detection.

The Carina region has been observed with several gamma-ray experiments. 
\textit{Agile} detected the source \object{1AGL J1043-5931} ($\alpha$=161.2, $\delta$=--59.7, with an error of $0.4\degree$) that could be related to $\eta$~Carinae \citep{Tavani+09}. A 2-day $\gamma$-ray flare has also been observed in October 2008, although its origin is unclear.
{\it Fermi/LAT} detections where reported in the {\it Fermi} bright source list \citep[BSL,][]{2009ApJS..183...46A} and in the first year catalog \citep[1FGL, ][]{2010ApJS..187..460A}. In the BSL, the source 0FGL~J1045.6-5937 lies at $\approx$~5.4~arcmin of $\eta$~Carinae and is therefore not associated with it.
In the 1FGL, the source 1FGL~J1045.2-5942 lies at 1.7~arcmin from $\eta$~Carinae which is again too far for a formal identification. A possible association with the open cluster Trumpler~16, which may hold some energetic young pulsars, was suggested \citep{2010ApJS..187..460A}.
In the TeV domain, no detection has been reported in the vicinity of $\eta$~Carinae.

In the following sections, we describe our detailed analysis of the {\it Fermi/LAT} data, and discuss the resulting multi-wavelength spectrum of $\eta$ Carinae in terms of electronic and hadronic accelerations.

\section{{\it Fermi/LAT} observations}

The large area telescope (LAT) aboard the {\it Fermi} spacecraft is an electron-positron pair conversion telescope sensitive to photon energies from 20~MeV to $>300$~GeV \citep{2009ApJ...697.1071A}. The LAT is made of 16 towers, each including a particle tracker and a calorimeter (allowing for the reconstruction of the direction and energy of the incident gamma-ray photons), and of an anti-coincidence shield rejecting the charged particle background.

The data are provided by the Fermi Science Support Center\footnote{\url{http://fermi.gfsc.nasa.gov/}}, which also delivers dedicated analysis software, called the {\it Fermi ScienceTools}\footnote{\url{http://fermi.gfsc.nasa.gov/ssc/data/analysis/software}}. The software modules allow to select events from an incoming region of the sky, to compute the corresponding live-time exposure map and the likelihood of an input sky model, and to derive significance maps, source spectra and light-curves.

\subsection{Data analysis}
We have used {\it Fermi/LAT} data accumulated in the Carina region during 21 months, from August 4, 2008 to April 3, 2010. The reconstructed gamma-ray photons were selected in a circular region centered on $\eta$ Carinae with a radius of $10^{\circ}$, belonging to the \emph{diffuse class}\footnote{\url{http://www-glast.slac.stanford.edu/software/IS/glast_lat_performance.htm}}, and with energies between 200~MeV and 100~GeV.
This upper limit is justified by the fact that at higher energies, systematics of the instrument and the role of cosmic-ray events are not yet fully understood.
The low energy threshold was chosen for several reasons.
Firstly, $\eta$~Carinae is located in the Galactic plane, where the galactic diffuse emission (whose spectrum can be described by a powerlaw with a photon index of 2.2) is very strong. Selecting a threshold of 200~MeV decreases the number of background events by a factor of 4.
Secondly, the galactic diffuse emission, whose knowledge is crucial to estimate the sky model parameters at low energies, is affected by large systematic uncertainties, influencing the likelihood analysis.
Thirdly, the size of the LAT point spread function decreases rapidly with increasing energy (from 100~MeV to 200~MeV, the $68 \%$ event containment radius decreases from $5^{\circ}$ to $3^{\circ}$). As the analysed region encompasses a large number of sources (25 sources are listed in the 1FGL within $10^{\circ}$ of $\eta$ Carinae, out of which 10 are within $3.5^{\circ}$), the ability to distinguish photons from different sources is important.

\subsection{Localisation}

\begin{figure}[h!]
 \centering
 \includegraphics[height=12cm]{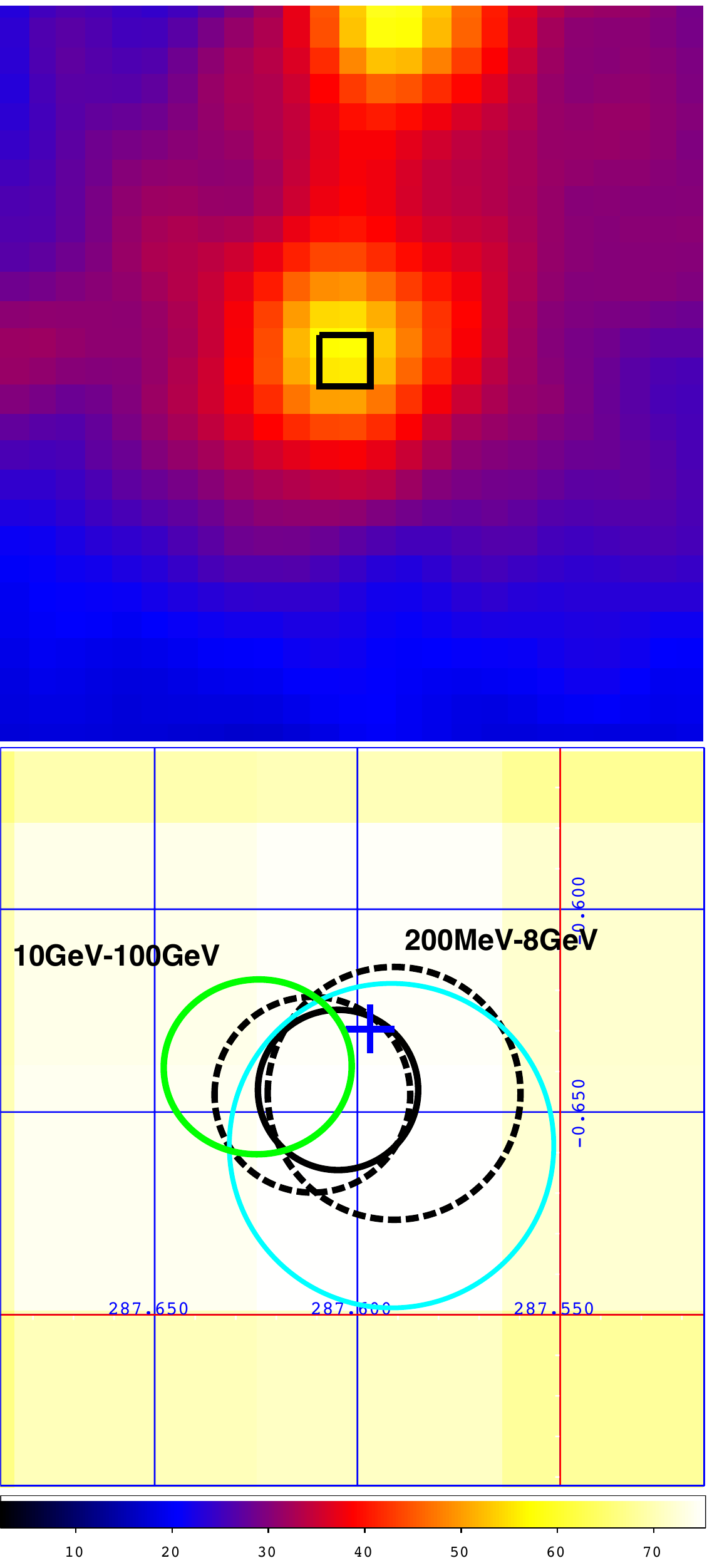}
 \caption[]
 {
   {\it Top:} 
   {\it Fermi/LAT} smoothed count map between 200~MeV and 100~GeV in galactic coordinates.
   The bright source in the centre of the field is FGL J1045.0-5942. The source located $\sim1.3^{\circ}$ to the North is the pulsar \object{PSR~J1048-5832}.
   {\it Bottom:}
   {\it Fermi/LAT} significance map corresponding to a zoom on the black box shown in the upper image.
   The blue cross indicates the optical location of $\eta$~Carinae.
   The green circle is the 95\% confidence region reported for 1FGL~J1045.2-5942 (based on 11 months of data).
   The plain black circle shows the 95\% confidence region found in our analysis of the 21 month dataset.
   The dashed circles show the 200~MeV to 8~GeV and 10~GeV to 100~GeV confidence regions.
   The cyan circle corresponds to the confidence region of the hard X-ray detection obtained with {\it INTEGRAL} \citep{Leyder2010}.
 }
 \label{fig:cntMap1D}
\end{figure}

The source \object{1FGL~J1045.2-5942} ($\alpha$=161.3053, $\delta$=--59.7057) was not formally associated with $\eta$ Carinae since the latter lies slightly outside of the $95 \%$ confidence region. As we are using almost twice more data than was available when the 1FGL was generated, a more accurate source localisation can be obtained. After an initial fit of the parameters of the sources close to the location of $\eta$~Carinae, as well as of the normalisations of the diffuse galactic and isotropic emissions, we determined the position of the nearest source to $\eta$~Carinae with the tool \texttt{gtfindsrc} and found $\alpha$=161.265 and $\delta$=--59.7015, with a $95\%$ confidence radius of 1.18~arcmin. $\eta$~Carinae is thus perfectly compatible with our improved position, as illustrated in Fig. \ref{fig:cntMap1D}. 

Since the width of the point spread function for photons converted in the front part of the LAT is smaller than for these converted in the back, we also derived the position using the front events only, but the uncertainty did not decrease significantly. We also derived the source location for low ($<$ 8 GeV) and high ($>$10 GeV) energy events and obtained error circles also compatible with the position of $\eta$ Carinae (see Fig.  \ref{fig:cntMap1D}).

The presence of the source is indubitable, with a test statistic \citep[TS,][]{0018.32003} $ > 2800$ ($\approx$ 53~$\sigma$) for the 200~MeV to 100~GeV energy range. The low and high-energy components have respectively TS of 2281 ($\approx$ 47~$\sigma$) and 73 ($\approx$ 8.5~$\sigma$).
In the rest of the paper, we will name the {\it Fermi/LAT} source \object{FGL J1045.0-5942}, owing to its improved position.

\subsection{Spectral analysis}
\label{subsec:spectrum}
The spectral analysis of FGL J1045.0-5942 was performed using the maximum likelihood method. The region modeling includes two diffuse emissions (Galactic plane and isotropic) and 35 point-like sources listed in the 1FGL catalogue. While most of these sources have been modeled as pure power-laws, four among them are known to be pulsars and have thus been modeled as power-laws with exponential cutoff \citep{2010ApJS..187..460A}. This is particularly important for the pulsar PSR~J1048-5832, located $\sim 1.3^{\circ}$ away from FGL~J1045.0-5942.

\begin{figure}[t]
 \centering
\hfill  \includegraphics[width=0.87\columnwidth]{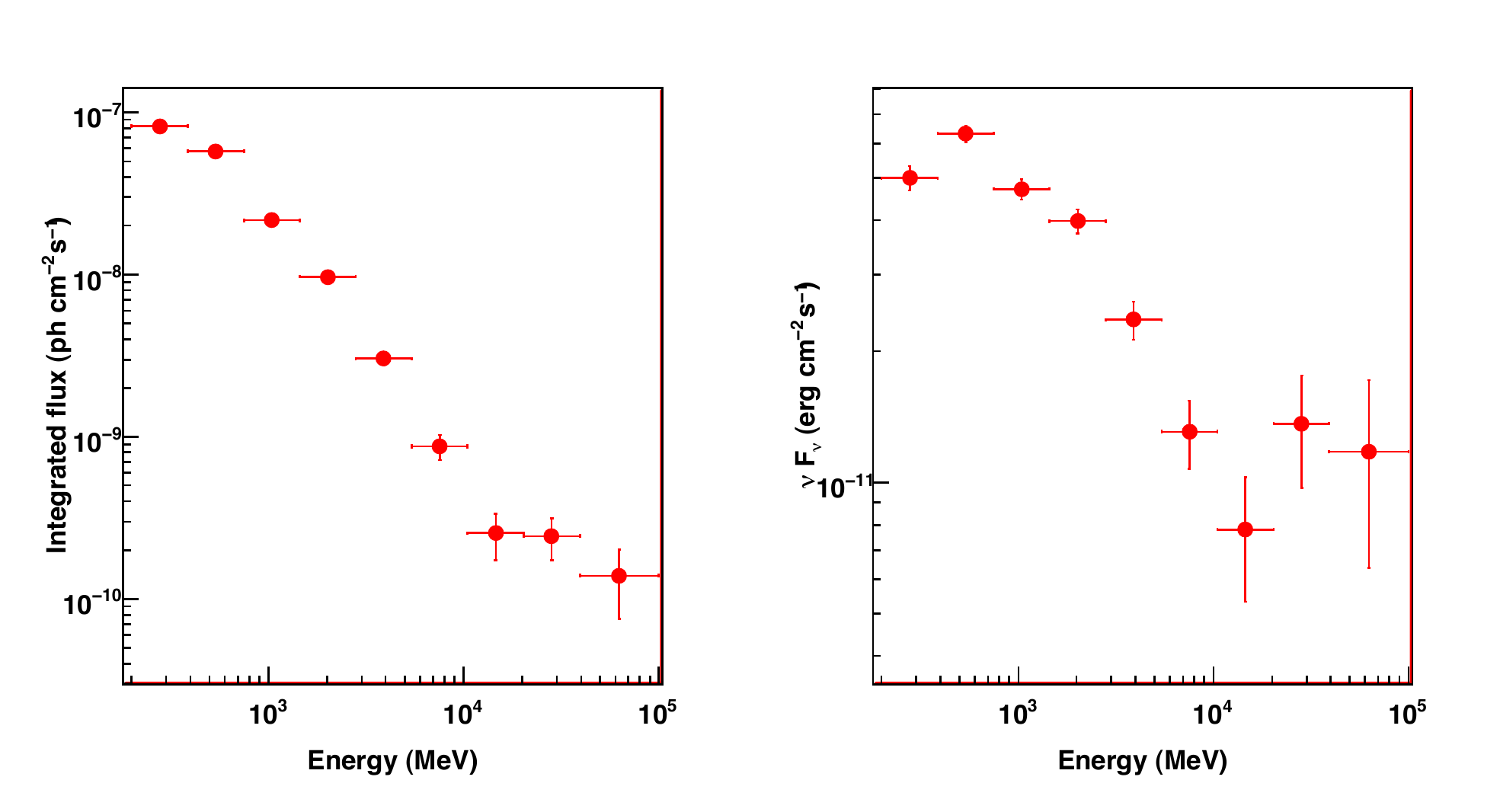}\\
\hfill  \includegraphics[width=0.9\columnwidth]{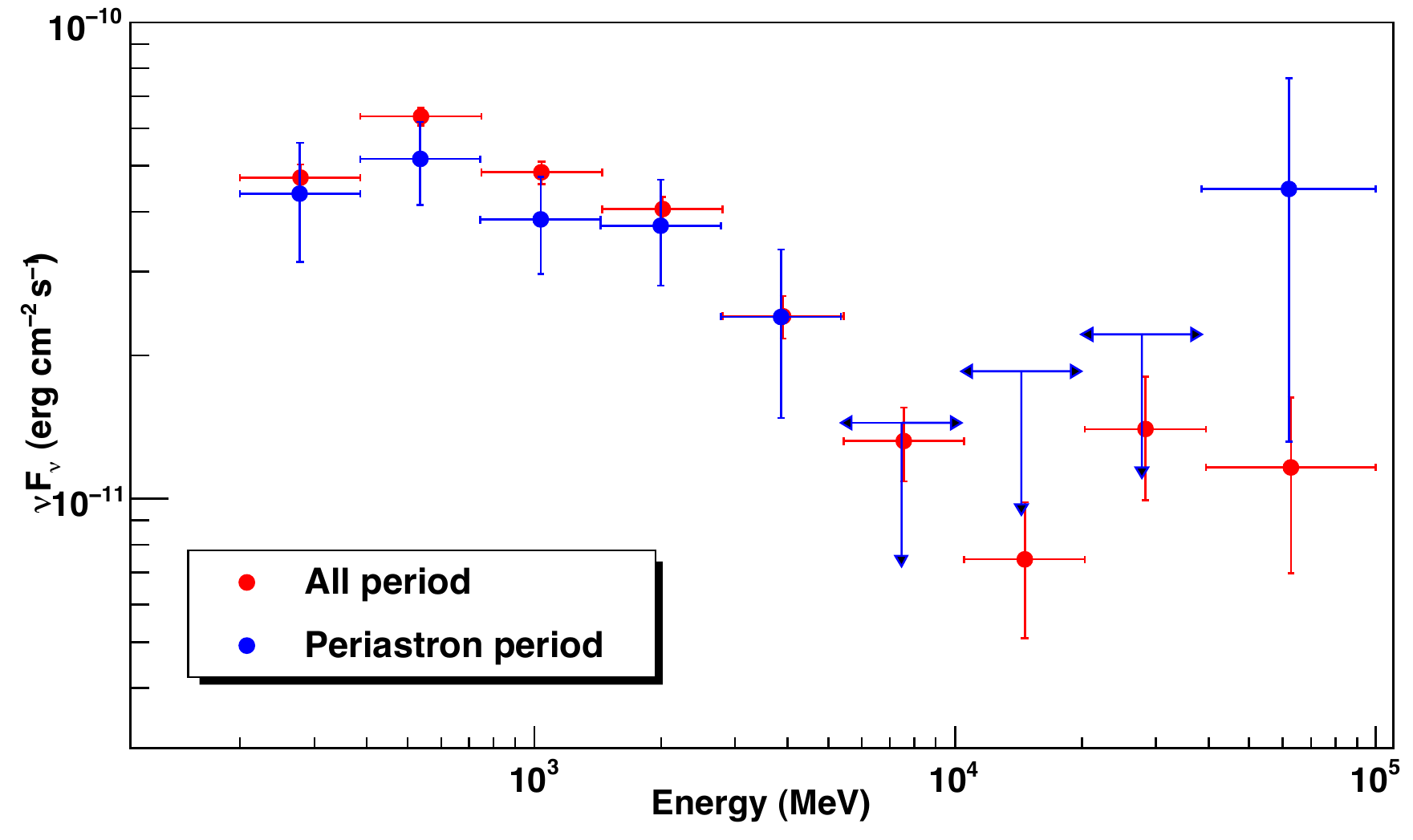}
 \caption[]
 {
{\it Top:}    Photon spectrum ({\it left}) and spectral energy distribution ({\it right}) of FGL J1045.0-5942 with 21~months of {\it Fermi/LAT} data.
{\it Bottom:} Spectral energy distribution for the the complete 21 month period (red) and for the periastron period (blue).
Uncertainties are 95\% confidence levels.
 }
 \label{fig:SED_FERMI}
\end{figure}

As the detection of FGL~J1045.0-5942 is very significant, a spectrum could be extracted in 9 spectral bins. In each bin, the TS is larger than 25 (5 $\sigma$). The resulting photon spectrum and spectral energy distribution are presented in Fig. \ref{fig:SED_FERMI}. Both binned  \citep{1979ApJ...228..939C} and un-binned \citep{EGRET_like} likelihoods have been used and their results are in full agreement. A curvature at low energy and a modification of the spectral slope at high-energy are clearly detected. 

The overall spectrum of FGL J1045.0-5942 differs from the standard power-law used to build the 1FGL catalogue, and has been modeled as a combination of two components: a power-law of photon index $\Gamma$ with an exponential cutoff plus another power-law at the highest energies.
Since only a limited number of spectral shapes are available in the {\it Fermi ScienceTools}, the dataset has been split into two energy ranges (0.2--8~GeV and 10--100~GeV) to determine the two spectral components independently. We maximized the sum of the likelihood in both spectral bands together so that the parameters of all sources are constrained by the complete dataset. The resulting parameters are $\Gamma = 1.69 \pm 0.12$, $E_{\rm cut} = 1.8 \pm 0.5$~GeV, F$_{\rm 0.2-100~GeV}=1.52\times 10^{-7} {\rm ~ph~cm^{-2}~s^{-1}}$ for the exponentially cutoff power-law and $\Gamma = 1.85 \pm 0.25$, F$_{\rm 0.2-100~GeV}=0.41\times 10^{-7} {\rm ~ph~cm^{-2}~s^{-1}}$ for the high-energy component.

\subsection{Timing analysis}

\begin{figure}[t]
 \centering
 \includegraphics[width=\columnwidth]{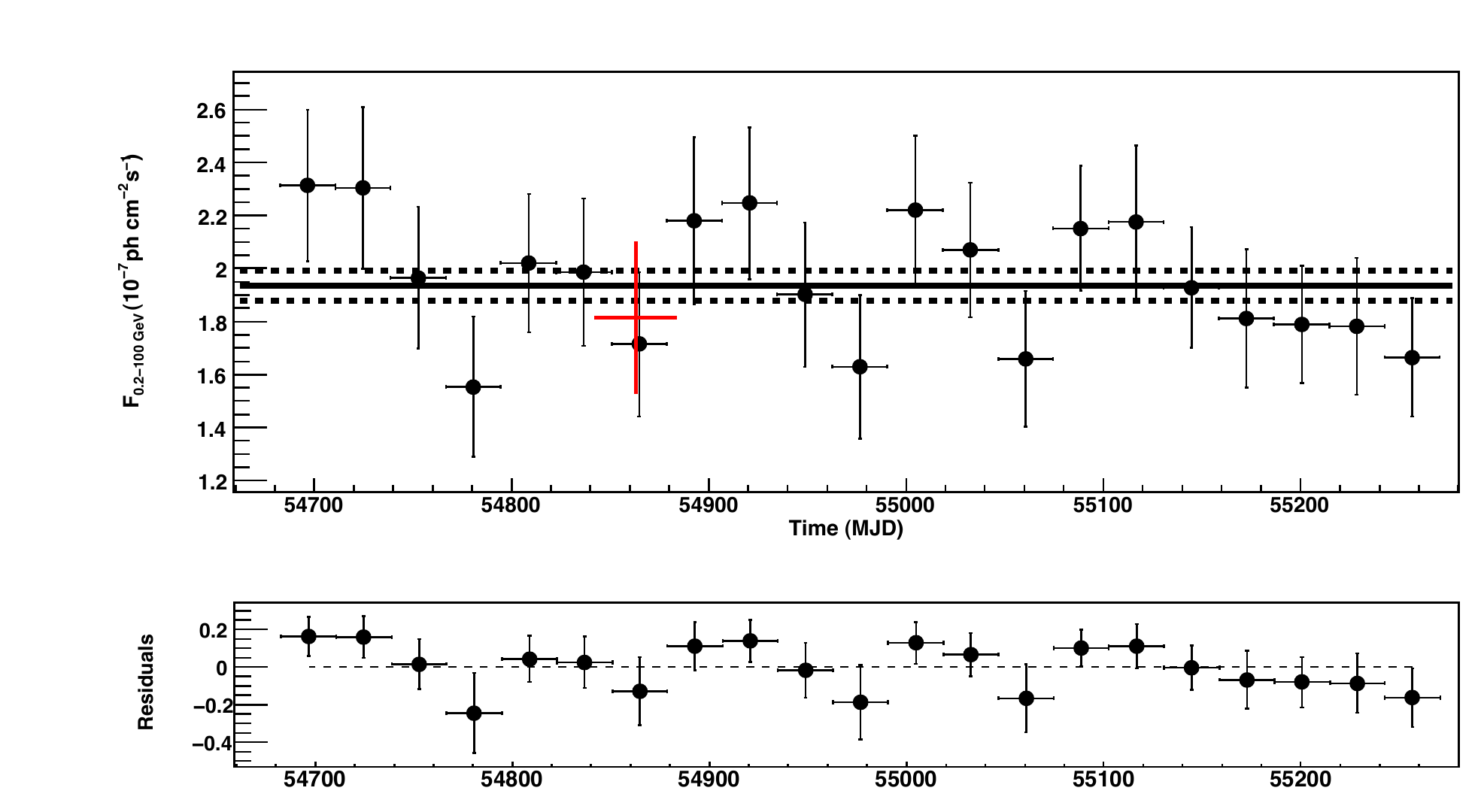}
 \caption[]
 {
   {\it Top: }
   Monthly light-curve of FGL J1045.0-5942 derived from the 21 month {\it Fermi/LAT} dataset. The red point corresponds to the flux level observed during the X-ray minimum.    
   The black line represent the best fit result (with its error) obtained assuming a steady flux for the source. Uncertainties are 95\% confidence levels.
   {\it Bottom:}
   Residuals obtained from the steady flux fit.
 }
 \label{fig:LC_FERMI}
\end{figure}

We constructed the light-curve of  FGL J1045.0-5942 using the maximum likelihood technique with time bins of one month (see Fig. \ref{fig:LC_FERMI}). 
The light-curve is consistent with a steady flux. The time-averaged flux of the source in the 0.2--100~GeV energy range is F$_{\rm 0.2-100~GeV} = (1.93 \pm 0.05) \times 10^{-7} {\rm ~ph~cm^{-2}~s^{-1}}$.

The flux was also determined for the period when $\eta$~Carinae was at the X-ray minimum, as measured by {\it RXTE}, (i.e. from MJD = 54843 to MJD = 54883). The flux level obtained for this period is reported in red in Fig. \ref{fig:LC_FERMI}. It is compatible with the average flux at the 95\% confidence level. The spectrum detected during the periastron period is also compatible with the average spectrum obtained over the 21 months period (bottom of Fig. \ref{fig:SED_FERMI}). In particular the cutoff energy did not change significantly, nor the high-energy component, clearly detected above 40 GeV.

The gamma-ray emission does not vary by more than 50\%, much less than observed in the X-ray band at periastron, where a variability factor larger than10 is observed. 

Fig. \ref{fig:RXTE_rate_and_high_energy_Fermi_events} shows the high-energy ($>$ 20 GeV) events superimposed on the {\it RXTE} X-ray light-curve\footnote{\url{ftp://legacy.gsfc.nasa.gov/xte/data/archive/ASMProducts/definitive_1dwell/lightcurves/}}. 
Despite the fact that this high-energy component is very significant ($TS>110, \sigma > 10$), there is also no apparent correlation between the high-energy component and 
the orbital phase of $\eta$ Carinae. 

We also searched for gamma-ray flares using a flux aperture photometry method, but no significant deviation has been found. In particular, the two-day period during which {\it Agile} reported a flare from the Carina region has been investigated, but no sign of activity could be detected. During these two days, the average flux detected by {\it Fermi/LAT} is $F = (1.9 \pm 0.6) \times 10^{-7} \mbox{ph cm}^{-2}\mbox{s}^{-1}$, well below the flux detected by {\it Agile} which is $(27.0 \pm 6.5) \times 10^{-7} \mbox{ph cm}^{-2}\mbox{s}^{-1}$ \citep{Tavani+09}. As {\it Fermi} and {\it Agile} have different orbits and as we used an energy threshold of 200~MeV, our analysis cannot exclude a very short and low energy flare.

\section{Discussion}
\subsection{Association with $\eta$ Carinae}

Since the spectral energy distribution observed by {\it Fermi/LAT} is made of two components (Fig. \ref{fig:SED_FERMI}), a spatial coincidence of two distinct gamma-ray sources (for instance a pulsar and a blazar) cannot be excluded a priori. However this coincidence should match with the centroids of the low- and high-energy components, which are separated by less than 1.2\arcmin\ (Fig. \ref{fig:cntMap1D}) and compatible within the uncertainties. This is rather unlikely as this would require two bright gamma-ray sources very close to $\eta$ Carinae. 

Given the number of pulsars detected within 2\degree\ of the Galactic plane (26 or 28\footnote{The 1FGL quotes two potential $\gamma$-ray millisecond pulsars within 2\degree\ of the Galactic plane.}), the probability to find one of them in the 95\% confidence region of $\eta$~Carinae is $2.5 \times 10^{-5}$. Moreover, we conducted a blind search frequency analysis and did not find any evidence for a pulsation in FGL J1045.0-5942.

\begin{figure}[t]
 \centering
   \includegraphics[width=0.95\columnwidth]{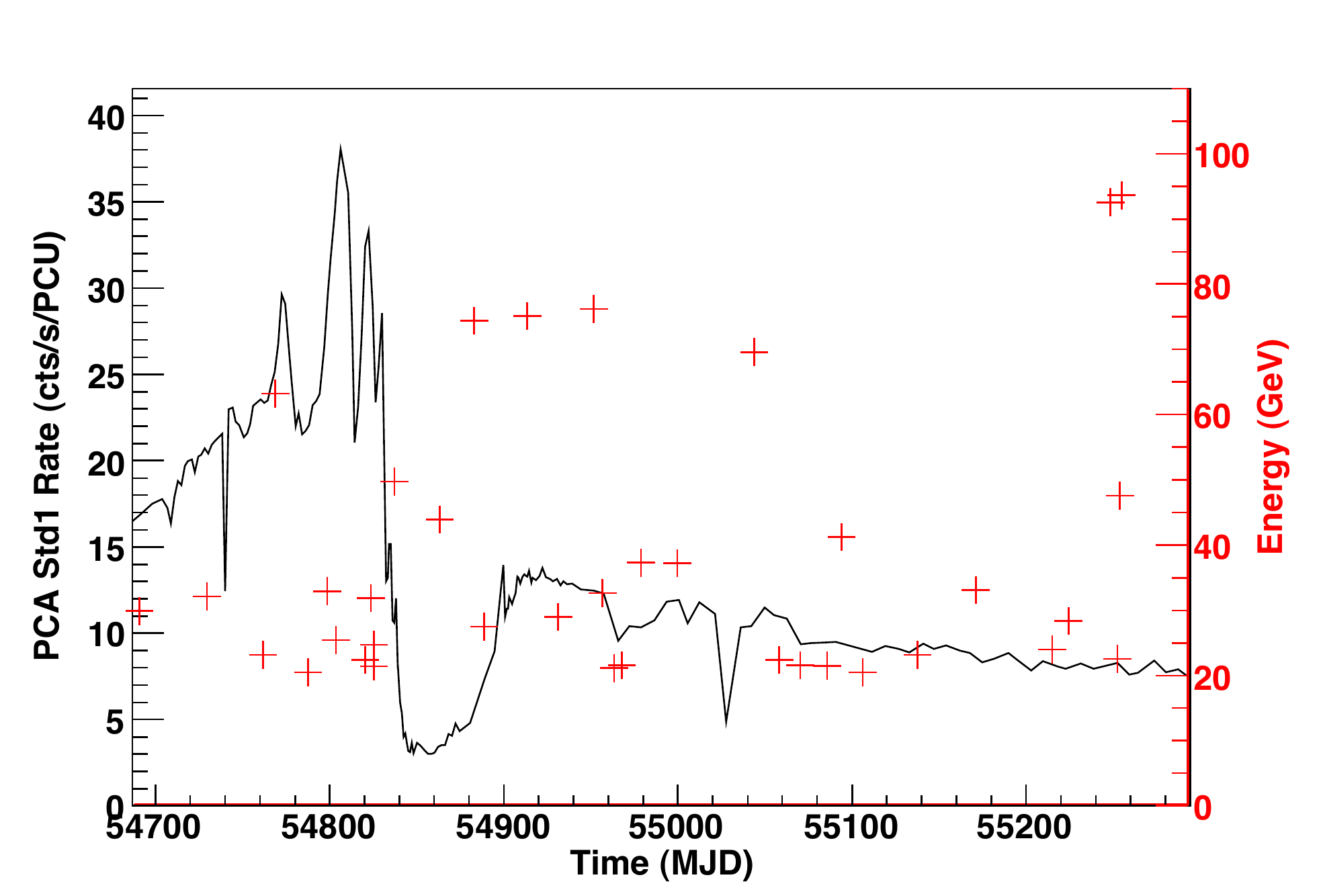}
 \caption[]
 {
   The black line shows the {\it RXTE} X-ray light-curve of $\eta$~Carinae with the minimum close to the 2009 periastron passage. The red points correspond to the high-energy events ($E > 20$~GeV) detected by {\it Fermi/LAT} within 1\degree\ of $\eta$ Carinae (note the event energy scale on the right).
}
 \label{fig:RXTE_rate_and_high_energy_Fermi_events}
\end{figure}

Given the 685 sources associated with blazars in the 1FGL over the entire sky, the probability to find a blazar at the location of $\eta$ Carinae is $2.3 \times 10^{-5}$.
In addition, if the high-energy emission was related to a blazar, flaring episodes would be expected, but none is observed (see Fig. \ref{fig:RXTE_rate_and_high_energy_Fermi_events}).

These two results imply that the probability to find a pulsar and a blazar at the location of $\eta$ Carinae is lower than $10^{-9}$.

With the above probabilities and the absence of any other bright X-ray source in the {\it INTEGRAL} error circle \citep{Leyder2010}, we can safely assume that the high-energy emission detected by {\it Fermi} and {\it INTEGRAL} comes from a single source, very likely $\eta$ Carinae.

{\it Chandra} images of $\eta$ Carinae \citep{2004A&A...415..595W} show an extended soft shell-like component (corresponding to an external shock at the boundary of the \textit{Homunculus} nebula) around the harder, point-like emission. \cite{2010arXiv1006.2464O} have argued that the gamma-ray emission detected by {\it Fermi} could be emitted by this external shock. Their conclusion was based on the first eleven month spectrum, which did not exhibit the two spectral components presented above. The high-energy spectral component cannot be explained by the external shock, because the density is far too low to allow significant hadron interactions and $\pi^0$ cooling. In addition, \cite{Leyder2010} have shown that the contribution of this outer shell to the hard X-ray emission from $\eta$ Carinae was lower than 15\%. Thus the outer shell interpretation seems unlikely.

\subsection{Particle acceleration in $\eta$ Carinae}

We will assume below that particle acceleration takes place in $\eta$ Carinae through diffusive shock acceleration in the colliding wind region of the binary system.
In the shock region, particle acceleration is counterbalanced by four main cooling processes:
\begin{enumerate}
\item {\bf Inverse Compton scattering} of electrons in the intense ultraviolet radiation field of the stars.  The cooling time scale is
$$
t_{IC} = \frac{3 \gamma m_e c^2} {4 \sigma_T c \gamma^2\beta^2U_{rad}}
= \frac{3 \pi R^2 m_e c^2}{\sigma_T \gamma\beta^2L} 
\approx \frac{R^{2}_{10^{14}\rm cm}}{\gamma_{10^4} L_{5\cdot10^6 \rm L_\odot}}\times 
6\cdot10^2~{\rm s},
$$
where $L$ is the stellar ultraviolet luminosity of the primary star and $R$ is its distance to the colliding wind region.
\\

\item {\bf Proton-proton interactions} and subsequent pion decay. \cite{2005MNRAS.363L..46B} pointed out that high-energy hadrons in the wind of massive stars will be photo-disintegrated into protons and neutrons on a short timescale. The interaction timescale for protons is inversely proportional to the density of matter in the post-shock region, where the particles are trapped by the magnetic field. Hydrodynamic simulations \citep{2009MNRAS.396.1743P} indicate that the density in the shock region is much larger than the unperturbed wind density, by a factor of $\delta\sim$1--100. The p-p cooling time scale can therefore be written :
$$t_{pp}= \frac{1} {\sigma_{pp} \delta n c}=\frac{4 \pi R_{sh}^2 m_p V_{w}} {\sigma_{pp} \delta \dot{M} c}\approx \frac{R^{2}_{10^{14}\rm cm} ~V_{10^{3}\rm km/s}}{\delta_{10}~\dot{M}_{10^{-4}\rm M_\odot /yr}}\times 4\cdot 10^5~ {\rm s},$$
where $V$ is the typical wind velocity and $\dot{M}$ is the mass loss rate of the primary star.
\\

\item {\bf Proton diffusion} away from the shock region. The diffusion timescale is related to the bulk velocity of the post-shock material from the central regions towards the outside. Hydrodynamic simulations indicate that the material flows out of the shock region at a velocity smaller than, but of the order of,  the pre-shock wind velocity \citep{2009MNRAS.396.1743P}. As the shock region has a size similar to the stellar separation, the bulk diffusion timescale can be estimated as (\citealt{2005MNRAS.363L..46B}) :
$$ t_{bulk}=3R/V\approx \frac{R_{10^{14}\rm cm}}{V_{10^3\rm km/s}}\times 3\cdot 10^6~ {\rm s} .$$

\item {\bf Electron bremsstrahlung} in the post-shock region. For a density 
$$\delta n= \delta \frac{\dot{M}}{4 \pi R^{2}V m_{p}} \approx \frac{\delta_{10}\dot{M}_{10^{-4}M\odot/yr}}{R_{10^{14}cm}}\times 3 \cdot 10^{9} {\rm cm^{-3}}, $$ 
the cooling timescale is given by  \citep{2004vhec.book.....A} :
$$t_{br} \approx 1.2 \times 10^{15} (\delta n_{\rm cm^{-3}})^{-1} {\rm s} \approx 4 \cdot 10^{5}~{\rm s}.$$
\end{enumerate}

Particle acceleration in the shock is therefore mainly counterbalanced by Inverse Compton scattering for electrons and by proton-proton interactions.
Equating the acceleration time $t_{acc}=\frac{R_L}{c}\left( \frac{c}{V}\right)^2$ respectively with $t_{IC}$ and $t_{pp}$ for electrons and protons ($R_L$ is the Larmor radius)
provides the maximum characteristic energy of the particle distribution. 

For electrons, a powerlaw spectrum is expected, with an exponential cutoff at an energy
$$\gamma_{max, e}=
\sqrt{\frac{3\pi e c^2}{\sigma_T \beta^2}} \sqrt{\frac{B\cdot R^2}{L}} \frac{V}{c}\approx \sqrt{\frac{B_{1\rm G}\cdot R_{10^{14}\rm cm}^2}{L_{5\cdot 10^6 \rm L_\odot}}}V_{10^3\rm km/s} \times 3\cdot 10^4,
$$
where $B$ is the magnetic field in the shock region.

$\gamma_{max, e}$ should be fairly independent of the orbital position in the dipole approximation for a magnetic field varying as $R^{-2}$. 
The gamma-ray spectrum will therefore feature an exponential cutoff at the maximal energy as derived by \cite{1993ApJ...402..271E}.
A cutoff energy of $\sim 1$ GeV (Sect. \ref{subsec:spectrum}) corresponds to a magnetic field of $\sim50~\rm G$ at the stellar surface. 

For protons, the maximum characteristic energy is limited by proton-proton interactions to 

$$\gamma_{max, p}=
\frac{4\pi R^2 e B}{\sigma_{pp}\delta \dot{M}}\left(\frac{V}{c}\right)^3\approx\frac
{R_{10^{14} \rm cm}^2 B_{1\rm G} V_{10^3~\rm km/s}^3}
{\delta_{10} \dot{M}_{10^{-4}\rm M_{\odot}/yr}}\times  4\cdot10^4.  
$$

\begin{figure}[t]
 \centering
 \includegraphics[width=\columnwidth]{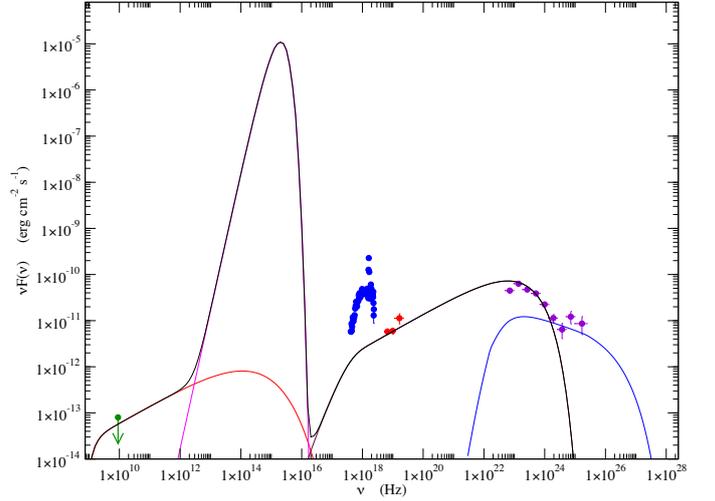}
 \caption[]
 {
   Spectral energy distribution of $\eta$~Carinae including {\it BeppoSAX/MECS} (blue), {\it INTEGRAL/ISGRI} (red) and {\it Fermi/LAT} (purple) data, and a radio upper limit to the synchrotron emission (green). 
   From low to high energies are shown the synchrotron, stellar emission, inverse Compton and $\pi^0$-decay spectral components.
 }
 \label{fig:multiple_SED}
\end{figure}

The non-thermal spectral energy distribution of $\eta$ Carinae has been represented with a model consisting of two cutoff powerlaw distributions for the electrons and for the interacting protons. The model and the data are shown together in Fig. \ref{fig:multiple_SED}. The parameters of the model are listed in Table \ref{tab:modelpar}.
The magnetic field and the electron energy distribution were adjusted to match the upper limit on the radio synchrotron emission, derived from the minimal thermal emission detected with {\it ACTA} \citep{2003MNRAS.338..425D}, and to match the inverse Compton continuum determined by {\it INTEGRAL} and {\it Fermi}. The slope and cutoff energy of the interacting proton energy distribution were fixed to 2.25 and $10^4$, respectively, and its normalization was fitted to match the high-energy gamma-ray tail using the $\pi^0$-decay model of \cite{2006PhRvD..74c4018K}. The proton and ultraviolet photon energy densities in the shock region were fixed to the values expected for an average distance $R=10^{14}$ cm.

\begin{table}[t]
\caption{Model parameters of the shock region determining the non-thermal spectral components of $\eta$ Carinae, as displayed in Fig. \ref{fig:multiple_SED}.}
\label{tab:modelpar}
\centerline{
\begin{tabular}{l|l|rl}
\hline \hline \noalign{\smallskip}
&Parameter & \multicolumn{2}{l}{Value} \\
\noalign{\smallskip\hrule\smallskip}
Environment &Photon energy density&2.7&erg/cm$^3$\\
&Magnetic field&0.5&G\\
&Density&$3\cdot 10^9$&cm$^{-3}$\\
\noalign{\smallskip\hrule\smallskip}
Electron distribution&Powerlaw index&2.25&\\
&$\gamma_{\rm max, e}$&$10^4$&\\
&Total energy&$10^{40}$&erg\\
\noalign{\smallskip\hrule\smallskip}
Proton distribution&Powerlaw index&2.25&\\
&$\gamma_{\rm max, p}$&$10^4$&\\
&Total energy&$1.3\cdot10^{40}$&erg\\
\noalign{\smallskip\hrule\smallskip}
\end{tabular}
}
\end{table}

\subsection{Energetic Considerations}

The wind momentum ratio of $\eta$ Carinae is $\eta=(\dot{M}_2 V_2) / (\dot{M}_1 V_1) \approx0.2$ \citep{2002A&A...383..636P}. The half-opening angle of the shock region is therefore $\sim 1$ rad, the fraction of the wind involved in the wind collision region is $\sim10\%$, and the mechanical energy available in that region is $\sim 200~{\rm L}_\odot$.

For a density of cold protons of $3\cdot 10^{9}$ cm$^{-3}$ in the shock region, the normalization of the $\pi^0$-decay spectrum (Fig. \ref{fig:multiple_SED}) requires a total interacting proton energy $E_{p}\sim 10^{40}$ erg. The energy injected in the shock to sustain the observed proton distribution is of the order of $E_p/t_{pp}  \sim 10 ~{\rm L_\odot}$.

The gamma-ray spectrum of $\eta$ Carinae thus indicates that $\sim5\%$ of the shock mechanical energy (or less than 1\% of the total wind mechanical luminosity) is transferred to accelerated protons downstream. This is in agreement with recent numerical simulations of relativistic collisionless shocks \citep{2008ApJ...682L...5S}, if a significant fraction of the hadrons interacts and generates $\pi^0$.

Modeling the thermal X-ray emission depends on detailed hydrodynamical simulations \citep[see e.g. ][]{2002A&A...383..636P} and could be affected by many phenomena \citep{2009MNRAS.394.1758P}. Such simulations can explain an X-ray luminosity of $\approx 20~{\rm L_\odot}$ above a few keV. Our analysis hence indicates that the fraction of the shock energy accelerating protons is similar to that emitting observable X-rays in $\eta$ Carinae.

The ratio between the inverse Compton or $\pi^0$-decay and the X-ray emission observed in $\eta$ Carinae is larger than predicted by existing models applied to WR 140 \citep{2006MNRAS.372..801P,2006ApJ...644.1118R}. The high-energy emissivity predicted for WR140 is however not well constrained and varies by a factor of 100 depending on the model parameters. The strong inverse Compton emission and the enhanced $\pi^0$-emission (when compared to the X-ray emission) might be related to the strong ultraviolet photon field and to the very high density in the wind collision region of $\eta$ Carinae, strengthening simultaneously the proton-proton collision rate and the absorption of the thermal emission. Detailed modeling is outside the scope of this paper, but these observations of $\eta$ Carinae provide the first observational constraint on the fraction of the mechanical power injected into particle acceleration.

\subsection{On the detection of neutrinos}

The high-energy gamma-rays observed in $\eta$ Carinae are likely produced by $\pi^0$-decay, a process which produces as many neutrinos as gamma-rays.
Thus, the detection of neutrinos would provide a conclusive evidence that hadronic acceleration is at work in $\eta$~Carinae.

The neutrino spectrum resulting from the decay of $\pi^0$ has been studied by \cite{2007JPhCS..60..243K}. In the case of $\eta$ Carinae, and
based on $\gamma_{max,p}$$\sim$10$^4$, the expected spectrum is
$$ \frac{dN_{\nu}}{dE_{\nu}} \approx \left ( \frac{E_{\nu}}{1~{\rm TeV}} \right )^{-2.35} {\rm e}^{- \sqrt{E_{\nu}/0.25~{\rm TeV}}}\times 10^{-12}~({\rm cm^2~s~TeV})^{-1}.$$

The atmospheric neutrino background is far too high to expect any detection at such a low energy. However, since we have not measured the high-energy cutoff of the $\pi^0$-decay emission, we cannot exclude that Cherenkov experiments might detect $\eta$ Carinae at energies higher ($\gamma_{\rm max, p} \gtrsim 10^5$) than anticipated. Unfortunately, even in that case, only a few neutrinos could be detected by {\it KM3Net} in 5 years of observations, not enough for a source detection.

\subsection{Other sources}

The density and magnetic field in the vicinity of the shock in $\eta$ Carinae are such that a large fraction of the energy carried by high-energy protons is converted to gamma-rays. Systems with lower stellar winds or magnetic fields will feature fainter $\pi^0$-decay emission. 

Scaling the emissivity of $\eta$ Carinae by the mass loss rate (i.e. the wind density) suggests that LBV systems could be detected by {\it Fermi} up to the Galactic center while WR or OB systems could be detected within 1 kpc and 0.1 kpc, respectively. Very few objects could therefore be detected. 
OB associations could have a combined stellar mass loss rate as strong as $\eta$ Carinae and similar efficiency in creating $\pi^0$ \citep{1997MNRAS.288..237O}. They could thus be detected up to distances of several kpc, although their extended emission is more difficult to measure, especially in the Galactic plane.

The mechanical energy in the stellar wind of a massive star, integrated over its lifetime, could reach some $10^{50}$ erg. From an energetic point of view, extrapolating on the particle acceleration efficiency of $\eta$ Carinae, stellar winds ejected during the massive star evolution in a suitable environment could be as effective as a SNR to accelerate hadrons. Future Cherenkov observations will tell if stellar wind collisions could accelerate particles up to to the knee of the cosmic-ray spectrum.

\section{Conclusions}

We have detected a bright gamma-ray source at the position of $\eta$ Carinae using 21 months of {\it Fermi/LAT} data. Its flux at a few 100 MeV corresponds very well to the extrapolation of the hard X-ray spectrum of $\eta$~Carinae (as measured by {\it INTEGRAL} and {\it Suzaku}) towards higher energies. 
The spectral energy distribution, which corresponds to an average over almost half of the orbit of $\eta$ Carinae, features two spectral components. 

The first one is a powerlaw extending from keV to GeV energies, with an exponential cutoff at $\sim 1$ GeV. This component can be understood 
assuming inverse Compton scattering of stellar photons by electrons accelerated up to $\gamma\sim 10^4$ in the wind collision region. 
The observed cutoff energy implies a magnetic field $\sim 50$ G at the stellar surface. 

The second component, a hard gamma-ray tail, is detected above 20 GeV. This bright component could be explained by $\pi^0$-decay of accelerated hadrons interacting with the dense stellar wind in the shock region. The ratio between the fluxes of the $\pi^0$ and inverse Compton components is roughly as predicted by simulations \citep{2006MNRAS.372..801P,2006ApJ...644.1118R}. Bremsstrahlung emission is expected at a much lower level and with a cutoff energy similar to that of the inverse Compton component.

The hard gamma-ray tail can only be understood if emitted close to the wind collision region. Indeed, the external shock between the \textit{Homunculus} nebula and the interstellar medium occurs at a density and magnetic field strength that are much too small to explain the observed emission. It is possible that a part of the soft gamma-ray emission comes also from inverse Compton scattering of infrared photons in the external shock as suggested by \cite{2010arXiv1006.2464O} \citep[see also][]{Leyder2010}. It is however difficult to account for the hard tail without a significant fraction of the soft gamma-ray emission being emitted by the colliding wind region. Detection of any hard X-ray or soft gamma-ray variability would rule out the external shock model.

The energy transmitted to the accelerated particles ($\sim 5\%$ of the wind collision mechanical energy) is of the same order as that of the observed thermal X-ray emission, providing an important observational constraint for future numerical hydrodynamical models of the colliding wind region in $\eta$ Carinae.

With the exception of the short flare detected by {\it Agile}, the averaged hard X-ray \citep{Leyder2010} and soft gamma-ray emissions do not vary by more than 50\% along the orbit,  not even during the periastron passage. This provides an interesting constraint on the evolution of the magnetic field and of the wind collision region along the orbit.

We finally suggest that Cherenkov telescopes could measure the hadronic cutoff energy in $\eta$ Carinae.

\bibliography{bib_etacar}
\bibliographystyle{aa}

\end{document}